\documentstyle[aps,eqsecnum,tighten,epsfig]{revtex}
\input epsf.tex  

\def\e{\epsilon}
\def\Pl{\ell_P}
\def\q{\hat{q}}
\def\U{{\cal U}}
\def\H{{\cal H}}
\def\be{\begin{equation}}
\def\ee{\end{equation}}
\def\g{{\gamma}}
\def\a{\alpha}
\def\b{\beta}
\def\k{\kappa}
\def\h{{\bf h}}
\def\A{{\cal A}}
\def\O{{\cal O}}
\def\D{{\cal D}}
\def\G{{\cal G}}
\def\Ab{\overline{\A}}
\def\Gb{\overline{\G}}
\def\Cyl{{\rm Cyl}}
\def\S{\Sigma}
\def\W{{\cal W}}
\def\T{\Cyl}
\def\la{{\bf a}}
\def\lg{{\bf g}}
\def\Tr{{\rm Tr}}
\def\TA{{\rm TA}}
\def\SG{{\rm SG}}
\def\GS{{\rm GS}}
\def\Diff{{\rm Diff}}
\def\av{{\rm av}}
\def\ba{\begin{eqnarray}}
\def\ea{\end{eqnarray}}
\def\N{{\cal N}}

\newtheorem{Theorem}{Theorem}[section]
\newtheorem{Lemma}{Lemma}[section]
\newtheorem{Definition}{Definition}[section]

\begin{document}

\title{Diffeomorphism invariant Quantum Field Theories of Connections in 
terms of webs.}
\author{J. Lewandowski\thanks{ Instytut Fizyki Teoretycznej, Uniwersytet 
Warszawski,  ul. Hoza 69, 00-681 Warszawa, Poland}, 
T. Thiemann\thanks{lewand@fuw.edu.pl, thiemann@aei-potsdam.mpg.de}} 
\address{Albert-Einstein-Institut,
MPI f. Gravitationsphysik, Schlaatzweg 1, 14473 Potsdam, Germany}
\date{\today}

\maketitle
\begin{abstract}
In the canonical quantization of gravity in terms of the Ashtekar
variables one uses paths in the 3-space to construct
the quantum states. Usually, one restricts oneself to   
families of paths admitting only finite number of isolated
intersections. This assumption implies a limitation
on the diffeomorphisms invariance of the introduced structures.
In this work, using the previous results of Baez and Sawin,
we extend the existing results to a theory admitting all 
the possible piecewise smooth finite paths and loops. In particular, we
$(i)$ characterize the spectrum of the Ashtekar-Isham configuration
space, $(ii)$ introduce spin-web states, a generalization of the
spin-network states, $(iii)$ extend the diffeomorphism averaging
to the spin-web states and  derive a large class of diffeomorphism invariant 
states and finally $(iv)$  extend the 3-geometry operators and the 
Hamiltonian operator.

\end{abstract}

\section{Introduction}
In the context of the canonical quantization of diffeomorphism invariant
theories of
connections for a compact gauge group, in particular, quantum gravity in
four spacetime dimensions,  the quantum configuration space 
is coordinatized by certain elementary operators which could be called
generalized Wilson functions. These functions are labelled in general 
by a set of paths in the manifold in question. It turns out that 
the precise nature of these paths has a deep impact on the structure
on the resulting quantum theory.  All the considerations 
in quantum gravity to date concern  the case
that the paths defining a given Wilson function
can intersect only finitely many times. To insure this, piecewise
analyticity was often assumed. Therefore  in the sequel we loosely denote
that case   by ``the analytic category". In ``the smooth category", however, 
the known results are sparse. 

In this letter we extend the pioneering work by Baez and Sawin on this
issue. Specifically, this letter is organized as follows :\\

In section 2 we state the problem with the extension of results from
the analytic to the smooth category and recall the results due to Baez and
Sawin. Then we prove a master theorem which extends the results due to
these authors and which is key to sections 3, 4 and 5. 

In section 3 we apply the master theorem to show how the results 
concerning the Ashtekar-Isham algebra can be immediately 
extended.

In section 4 and 5 we define the notion of a spin web in analogy to the
spin-nets of the analytic category in order to apply the master theorem  
to the extension of results by Ashtekar, Marolf, Mour\~ao and the
authors concerning diffeomorphism invariant states of connections and
associated diffeomorphism invariant operators.

Finally, in section 6 we apply the machinery of the previous sections to
the operators that have proved to be elementary building blocks for the 
quantization of  gravitational field.

\section{The algebra of cylindrical functions on the space of connections}
\label{s1}

This section is divided into three parts. In the fist part we recall the
basic notions of cylindrical functions on the space of distributional
connections and inductive limit structures and review the results 
that one obtains if one labels cylindrical functions by piecewise
analytic, finitely generated graphs. In the second part we review the
construction, due to
Baez and Sawin, of a new label set, called ``webs", which are appropriate
for the case that the loops defining the Wilson functions 
are only piecewise smooth and may intersect each other an infinite number 
of times. Finally, in the third part we prove the 
master theorem on which the rest of this letter is based.

\subsection{Preliminaries}
\label{s1.1}

In the loop quantization of gravity as initiated by Ashtekar \cite{AA} 
and Jacobson, Rovelli and Smolin \cite{JRS} and Gambini et. al. (see
e.g. \cite{GP} and references therein) through
embedding general relativity into a Yang-Mills theory phase space and 
making use of Wilson loop variables as well as 
in a general framework for quantization of a diffeomorphism 
invariant theory of connections
with local degrees of freedom \cite{ALMMT2}
one uses the algebra of cylindrical functions defined
by the parallel transport maps as the main device [4-30]. The algebra, 
let us call 
it $\Cyl$, is a sub-algebra of the algebra of functions  defined 
on a space of $G$ connections $\A$.  The connections are defined on a bundle 
over a manifold $\S$. For the simplicity we assume here that the bundle 
is trivial \footnote{Most of the results concerning the gauge
invariant cylindrical functions can be easily
generalized to a non-trivial bundle case \cite{B1,AL1,AL2}.}
and we fix a global gauge (section).
We are assuming that $G$ is compact and semi-simple.
A  complex valued function $\Phi : \A \rightarrow  {\bf C}$
is called cylindrical, if there are piecewise smooth  
paths $p_1,...,p_n$\footnote{By a piecewise smooth path we mean here
a piecewise smooth map $p: [t_0,t_1]\ \rightarrow\ \Sigma$
continues on the whole interval of $[t_0,t_1]$
such that its (one-sided) derivative never vanishes except for the 
constant path; here we are actually interested in equivalence classes of 
paths where two paths are equivalent iff they differ by
an orientation preserving piecewise smooth diffeomorphism
$[t_0,t_1]\mapsto [s_0,s_1]$.} 
in $\Sigma$ and a function $\phi\in C^0(G^n)$  such that 
\be\label{cyl}
\Phi(A) = \phi(U_{p_1}(A),...,U_{p_n}(A))
\ee
where $U_p(A)\in G$  represents the parallel transport along the path $p$  
with respect to the connection $A$. 
To control the algebra it is useful to  decompose $\Cyl$ into a union
\be  \label{union}
\Cyl\ =\ \cup _{w\in \W}\Cyl_w
\ee
of  sub-algebras that are easier to handle ($\W$ is a labeling set
and will be specified below). This decomposition
is particularly convenient if it defines a so called `inductive
limit', that is, if for every $w_1, w_2\in \W$, there exists $w'\in \W$ 
such that 
\be
\Cyl_{w_1},\Cyl_{w_2}\subset \Cyl_{w'}.
\ee  
One might take for $\W$ the set of all the finite families
of piecewise smooth paths in $\S$ and associate
to each finite family of paths $w$  the sub-algebra  $\Cyl_w\subset \Cyl$
given by all the paths obtained from the elements of $w$ and their inverses
by using the path product.  
 That does define an inductive limit decomposition, however, given a general
family of paths $w$, we do not know much about the corresponding 
$\Cyl_w$. An example of a  sub-algebra $\Cyl_w$ we do know much about is the 
sub-algebra corresponding to a family of paths $w$ which is an embedded 
graph
\footnote{A graph is a {\it finite} family of 1-dimensional, oriented
sub-manifolds 
of $\Sigma$ such that every two can share only one or two points
of the boundary. Given a graph, its elements are called edges.
By an edge, we shall also mean an oriented 1-sub-manifold with
boundary.}. 
Then, $\Cyl_w$ can be identified with the algebra of functions   
$C^0(G^n)$ where $n$ equals the number of edges in $w$, via the map
\be\label{proj} 
\A \ni A\ \mapsto\ (U_{e_1}(A),...,U_{e_n}(A))\in \A_w\subset G^n.
\ee
where $e_i$ are the edges of $w$ and $\A_w$ is the image of that map. 
Such a map is defined for any
family of paths, but in the case of a graph  the map (\ref{proj}) is 
{\it onto}, therefore
\be
\A_w\ =\ G^n.
\ee
 If one fixes an analytic structure on $\S$ 
and defines $\Cyl_{(\omega)}\subset\Cyl$ to be the sub-algebra  
given by the piecewise analytic paths, then the analytic graphs
are enough to decompose $\Cyl_{(\omega)}$ into the following 
inductive limit sum
\be\label{unan}
\cup_{\g} \T_\g = \T_{(\omega)},
\ee
where the sum ranges over the set of piecewise analytic graphs in $\Sigma$.
That is true due to the following fact. Given two analytic graphs 
$\g,\g'$ there is an analytic graph $\g''$ such that $\g''\ge \g,\g'$, 
where  the inequality relation is defined as follows. Two families 
$w,w'$ of paths, are in the relation
\be \label{le}
w\ \le\ w'
\ee 
if every path-element of $w$ can be obtained from the elements of
$w'$ and their inverses by using the path product\footnote{Given
two (unparametrized) paths $p_1$ and $p_2$ such that the end
point of $p_1$ is a starting point of $p_2$, the path  product 
$p_2\circ p_1$ is a path obtained by connecting the end of $p_1$ 
with the beginning point of $p_2$.}.
Therefore, because of the composition rule of the parallel transports 
\be
U_{p\circ p'} = U_{p}U_{p'}
\ee 
we have
\be
\Cyl_{\g_1},\Cyl_{\g_2}\subset \Cyl_{\g'}.
\ee
The above decomposition was key to the following issues :
\begin{itemize}
\item The development of integral calculus on $\Cyl_{(\omega)}$ 
\cite{AL1,B1,MM}.
\item The characterization of the Gel'fand spectrum 
of the corresponding $C^*$ algebra when $G$ is compact \cite{AI,AL1,MM,AL3}.
\item Measure theory on the spectrum \cite{AL1,MM,AL3}.
\item The introduction of differential calculus on the spectrum
useful for introducing and studying the operators of quantum gravity 
\cite{AL2}.
\item The introduction of spin-networks as the orthonormal
basis with respect to the natural integral in  $\Cyl_{(\omega)}$ 
\cite{RS1,B2}.
\end{itemize} 
Our goal now is to see to what extent the above applications
can be generalized to the case of smooth paths. 

\subsection{The Baez-Sawin framework}
\label{s1.2}

The virtue of piecewise analytical graphs is that their edges cannot
intersect or overlap in an infinite number of distinct, isolated
points or segments. This is key to the definition of the algebra 
$\Cyl_{(\omega)}$ as an inductive limit since 
the union of two piecewise analytic graphs (after a finite number
of subdivisions) again defines a piecewise  analytic graph. This is no 
longer true when two graphs are just piecewise smooth, an example 
being given by the graph that consists of the union
of two smooth curves which intersect in a Cantor set \cite{RSI} or the 
intersection of the graph of the function $[0,1]\ni x\to
e^{-1/x^2}sin(1/x)$ with the $x$-axis. 
The graphs are not both contained in a same graph. 
Thus, for the full algebra $\Cyl$ the graphs are not enough
to provide an appropriate decomposition. 

Recently Baez and Sawin \cite{BS} introduced a suitable generalization
of an embedded graph which they called a {\it web}.

For the purpose of this paper, it is enough to know the following 
properties of the webs due to by Baez and Sawin:

$(a)$  a web is a finite family of piecewise smooth paths $e$
which do not self intersect;

$(b)$ for every web $w=\{e_1,...,e_n\}$ the image of the map (\ref{proj})
is a Lie subgroup of $G^n$;

$(c)$ for every finite family $w$ of smooth paths there exists a web
$w'$ such that $w\le w'$; 

Thus, in contrast to the picewise analytic graphs which are labelled
by its edges (i.e. paths which intersect at most at their endpoints) a
Baez-Sawin web is labelled by the paths $e_1,..e_n$ which may overlap or 
intersect each other possibly
an infinite number of times. Abusing the notation, we shall nevertheless
denote the labels of a web by edges again to distinguish them from 
an arbitrary path. Henceforth, we denote by $\W$ the set of all the webs 
in $\S$. We should also notice that Baez-Sawin's $\W$ is preserved
by the diffeomorphisms of $\S$. The highly non-trivial result proved 
in \cite{BS} is that $\W$ indeed {\it exists}. 

Due to property $(c)$, 
\be
\Cyl\ =\ \cup_{w\in\W}\Cyl_w,
\ee 
and $(\Cyl_w)_{w\in\W}$ is a projective family. The point $(b)$ 
gives us certain control on $\Cyl_w$. Our first result will
tell us more about $\Cyl_w$. 
\medskip

We end this section with  the Baez-Sawin definition of a web. For more  
details see \cite{BS}.    
A {\it web} is the (set) union of a finite number of 
families of edges  so-called {\it tassels}. First, to give the reader an 
intuitive picture, we explain  how  a graph gives rise to a web. Given a 
graph $\g$ 
subdivide each of its edges $e$ into two edges oriented in such a way that 
they end in the subdivision point. Denote the resulting graph by $w$.
A vertex of $w$ is either a vertex of $\g$ or one of the
subdivision points. Let $v$ be a vertex of $w$ which is also a vertex of 
$\g$. The set of edges of $w$ leaving $v$ forms a tassel
based at $v$ and the subdivided graph $w$ is a web (a special case of a web).

In general, edges of a tassel are allowed to intersect at points 
different from their end points and can overlap. Given a family
$T$ of edges, a point $q\in R(T)$\footnote{The {\it range} $R(T)$ of a 
family of curves $T$ is the union of the ranges of each element of $T$.}
is a {\it regular point} of $T$ if there is a neighborhood
${\cal U}$ of $q$ such that the intersection $R(T)\cap {\cal U}$
is an embedding of an open interval. A segment of that 
embedded interval we will call a {\it regular segment}.  
 (In other words, if a curve $e\in T$
intersects $q$ then it intersects $q$ exactly once, and any other
$e'\in T$ either  overlaps $e$ at $q$ or does not intersect $q$ at all.)
Given a regular point $q$ of $T$ its {\it type} is the set  of the elements
of $T$ which overlap at $q$.  
\medskip

\noindent{\bf General definition of a web.} A a family of edges is 
a {\it tassel}  if it  has the following  properties:

$(i)$ All the edges of $T$ begin at a same point; we call it
the {\it base point} of $T$;

$(ii)$ The edges $e_I$, $I=1,...,k$, of $T$ can be parameterized
in such a way that
\be
e_I(t)\ =\ e_J(s)\ \Rightarrow\ t=s;
\ee

$(iii)$ Two edges of $T$ that intersect at a point
different from the base point $p$ intersect in every 
neighborhood of $p$;    

$(iv)$ Any type which occurs at some point $q\in R(T)$ 
occurs in every neighborhood of the base point $p$.

%
%
%

\subsection{Holonomic independence of the curves in a Baez-Sawin web.}
\label{s1.3}

Paths $p_1,...,p_n$ are {\it holonomically independent}
if for every $(g_1,...,g_n)\in G^n$ there is a smooth connection
$A\in {\cal A}$ such that 
\be
(U_{p_1}(A),...,U_{p_n}(A))\ =\ (g_1,...,g_n)\in G^n.
\ee
As we already indicated in section (\ref{s1}),
the edges of any graph $\gamma$ are holonomically
independent, hence the corresponding $\Cyl_\gamma$ can be 
identified with the whole of $C^0(G^n)$. However, the results of \cite{BS}
only show that for an $n$-element web $w$ the set $\Cyl_w$ can
be identified with $C^0(G_w)$ where a priori $G_w$ can be any Lie
subgroup of $G^n$ that may even vary with $w$.  

Our master result is that it continues to hold that $G_w=G^n$ for the 
case that $G$ is semi-simple.
\begin{Theorem}[Master Theorem] \label{th1}
Let $\{p_1,...,p_n\}$ be a finite family of paths 
which satisfies the properties $(a)$ and $(b)$ above. Then
the paths are holonomically independent.
\end{Theorem}
{\bf Proof :} \\
For $n=1$ the assertion is obvious. For $n=2$
we need to notice that since $e_1\not=e_2$, there is a point 
$x\in R(e_1)$ such that a neighborhood $\U_1$ of $x$ does not
intersect $e_2$ and a similar point $y\in R(e_2)$ and a neighborhood
$\U_2$. The neighborhoods may be chosen not to intersect
each other. Therefore, given any $(g_1,g_2)\in G^2$
one can easily  construct a single connection $A$ such that 
$U_{e_1}(A)=g_1$ and  $U_{e_2}(A)=g_2$. 

Suppose now that the theorem is true for every $n\le k$ and consider the 
case when $n=k+1\ge 3$. Denote by ${\la}_w$ the Lie sub-algebra
of the Lie algebra of $G^n$ corresponding to the subgroup 
$\A_w=G_w\subset G^n$ obtained as the image of 
the map
\be \label{projl}
\A\ni\ \mapsto\ (U_{e_1}(A),...,U_{e_n}(A))\in G^n.
\ee
It suffices to prove that $ {\la}_w$ coincides with the whole
Lie algebra of $G^n$. Since, according to our assumption,
the theorem holds for $w' = w\setminus \{e_i\}$ for any edge $e_i$,
it is enough to show that for every $a\in {\lg}$ in the Lie algebra of
$G$ the $n$-tuple $(a,0,...,0)$ is in ${\la}_w$ since ${\la}_w$ is a
vector space (owing to Baez and Sawin).

Since $G$ is semi-simple, for any $a\in{\lg}$ there are $b,b'\in {\lg}$
such that $[b,b']=a$. By induction assumption, we can freely specify
$n$ of the entries $b_i$ of an element $(b_1,..,b_{n+1})$ of the Lie
algebra ${\la}_w$, however, the last entry may depend on the other $n$
ones. Let us choose freely $b_1=b,b_3=..=b_{n+1}=0$ and  
$b'_1=b',b'_2=b_4=..=b'_{n+1}=0$. Then there exist certain 
$\tilde{b},\tilde{b}'\in {\lg}$ which may depend on the already specified
$b_i$ and $b'_i$ respectively such that 
\be
(b,\tilde{b},0,...,0), (b',0,\tilde{b}',0,...,0)\in {\la}_w.
\ee 
Because ${\la}_w$ is closed with respect to the commutator
(again owing to Baez and Sawin), we have
\be
[(b,\tilde{b},0,...,0),(b',0,\tilde{b}',0,...,0)]\
=\ (a,0,...,0)\in {\la}_w,
\ee 
which completes the proof.\\
$\Box$

\medskip
 
\section{Immediate implications of the Master Theorem}

In this section we list the immediate consequences of theorem \ref{th1}
concerning the issue of (Mandelstam) group identities which are an
important 
building block in the construction of the Ashtekar-Isham algebra
\cite{AI}, the spectrum of that algebra and a natural uniform measure
thereon.

%

\subsection{Group Identities}
\label{s2.1}

The issue is the following :\\
Suppose that $k$ complex valued functions $\phi_1,...,\phi_k$ defined on
$G^n$ satisfy a group identity 
\be \label{id}
P(\phi_1,...,\phi_k)\ =\ 0\; \forall (g_1,..,g_n)\in G^n,
\ee
where $P$ is a complex valued function of $k$ complex variables
which is characteristic for the group $G$. Examples of such identities are
the famous Mandelstam identities which for the group $G=SU(2)$ are given
by 
\be \label{su(2)}
\mbox{tr}(g)\mbox{tr}(g')=\mbox{tr}(g g')+\mbox{tr}(g (g')^{-1}).
\ee
Now take any $n$-tuple
of paths $p_1,...,p_n$ in $\S$ and consider the following
cylindrical functions defined on $\Ab$
\be
\Phi_i(A)\ =\ \phi_i(U_{p_1}(A),...,U_{p_n}(A)), \ \ \ A\in \Ab, \ \ 
i=1,...,k.
\ee
Obviously these cylindrical functions satisfy then the identity 
\be
P(\Phi_1,...,\Phi_k)\ =\ 0 \;\forall A\in \A.
\ee 
Thus, we see that every group identity gives rise to {\it one} identity
on the algebra $\Cyl$.\\
The question that arises is whether {\it every} identity that holds on the
algebra $\Cyl$ (or one of its sub-algebras) declines from a group identity.

Owing to the master theorem, the answer turns out to be affirmative
(one could not prove this result without knowing theorem \ref{th1}).

Indeed, consider any system of cylindrical functions
$\Phi_1,...,\Phi_k\in \Cyl$ satisfying some identity
\be
P(\Phi_1, ... ,\Phi_k)\ =\ 0. 
\ee     
Any of the cylindrical functions is defined by (\ref{cyl})
with respect to some finite set of paths in $\S$. There exists a web 
$w=\{p_1,..,p_n\}$ whose range 
contains all the paths used to define the functions
$\Phi_1,...,\Phi_k$.  Thus $\Phi_i=p_w^\ast\phi_i$ for some functions
$\phi_i$ defined on the image $\A_w=G_w=p_w(A)$ of (\ref{projl}).
Consequently, the functions $\phi_i$ satisfy the corresponding identity 
(\ref{id}) on all of $G_w$. But according to theorem \ref{th1}
we have $G_w=G^n$ where $n$ is the number of the edges of $w$. Thus 
the identity on cylindrical functions came from a group identity.
Notice that, in particular, the path identities on cylindrical functions
\be
U_p(A)U_q(A)\ =\ U_{p\circ q}(A), \ \ \ , U_{p\circ p^{-1}}=1_G
\ee 
for any two paths $p,q$ and the Mandelstam identities, which for $SU(2)$
take the form
\be
\Tr U_{\a\circ 
\b} + \Tr U_{\a\circ\b^{-1}}\ =\  \Tr U_{\a} \Tr U_{\b}  ,
\ee
decline from a group identity.\\

A related question is : When are two paths $p$ and $q$ holonomically
equivalent. That is, for which paths $p$ and $q$ does the identity
\be
U_p(A)\ =\ U_q(A)
\ee 
hold on $\cal A$ (i.e. for all $A\in {\cal A}$) ? 

Let paths $p$ and $q$ be  holonomically equivalent.
Choose a web $w=\{p_1,...,p_n\}$ such that $\{p,q\}\le w$.
Decompose $p,q$ into the path products of the elements of $w$
and their inverses:  
\ba
p\ =\ e_{i(1)}^{\k(1)}\circ...\circ e_{i(N)}^{\k(N)}\nonumber\\
q\ =\ e_{j(1)}^{\l(1)}\circ...\circ e_{j(M)}^{\l(M)}.
\ea
Suppose, that in this decomposition, all the
sequences of the form $e_I\circ e_I^{-1}$ have been cancelled. 
 
The properties of the parallel transport,
the holonomic equivalence of the paths gives an identity
\be
g_{i(1)}^{\k(1)} ... g_{i(N)}^{\k(N)}\ =\ 
g_{j(1)}^{\l(1)}... g_{j(M)}^{\l(M)}
\ee
true for {\it every} $(g_1,...,g_n)\in G_w$. 

However, precisely because theorem \ref{th1} holds, $G_w=G^n$ which 
enables us, at least when $G$ is non-Abelian and compact, to 
to conclude that modulo the cancellation of  pairs $g_i g^{-1}_i$, 
\be
N\ =\ M,\ \ i(r) = j(r),\ \ \k(r)\ =\l(r). 
\ee  
That is {\it for a non-abelian compact Lie group, paths $p$ and $q$ 
are holonomically equivalent iff  $p=q$ modulo cancellation of the segments 
of the form $...\circ e\circ e^{-1}\circ...$.} (We are here relying 
on a related fact from the theory of compact non-Abelian groups 
\cite{result}). Notice 
that we could not conclude this 
if $G_w$ was a proper sub Lie group of
$G^n$.

\subsection{Spectrum of the Ashtekar -- Isham algebra}
\label{s2.2}

The main motivation for raising the question about the precise image 
of the map (\ref{proj}) was the wish to give a characterization of the 
Ashtekar-Isham quantum configuration space which in the analytical
category turned out to be rather straightforward. Recall, that when $G$ is
compact,
we define in the algebra of cylindrical functions $\Cyl$
a norm 
\be
\|\Phi\|\ :=\ \sup_{A\in\A}|\Phi(A)|.
\ee
The completion of $\Cyl$ with respect to that norm 
is a $C^*$ algebra. It's Gel'fand spectrum $\overline{{\cal A}}$ has
been promoted by Ashtekar and Isham to the role
of the quantum configuration space for a connection
theory.  According to the results of  \cite{MM,AL1,AL2,AL3}    
an element ${\bar A}$ of the spectrum of $\Cyl$ can be naturally 
identified with a family of points $({\bar A}_w)_{w\in\W}$ which satisfies
certain consistency condition. Namely, given two webs $w\le w'$,
the projections $p_w:\A\rightarrow \A_w$ and  $p_{w'}:\A\rightarrow 
\A_{w'}$ defined in (\ref{proj}) determine uniquely a projection
\be
p_{ww'} :\A_{w'}\ \rightarrow\ \A_w, \ \ {\rm s.t.}\ \ 
p_{ww'}\circ p_{w'}\ =\ p_w.
\ee 
The consistency condition satisfied by $({\bar A}_w)_{w\in\W}$  is that for 
every pair of webs $w\le w'$,
\be\label{con}
p_{ww'} {\bar A}_{w'}\ =\ {\bar A}_w.
\ee
More precisely, there is a 1-to-1 correspondence between the set of
solutions
of the consistency condition and the spectrum. To solve the
consistency conditions one considers a map $U({\bar A})$ which
assigns to every path $p$ an element of $G$,
\be
p\mapsto U({\bar A})_p\in G, 
\ee 
such that
\be
 U({\bar A})_{p\circ q}\ =\ U({\bar A})_p U({\bar A})_q,\ \ {\rm and}\ \
  U({\bar A})_{p^{-1}}\ =\  (U({\bar A})_p)^{-1}.   
\ee
Indeed,  a map  $U({\bar A})$ defines a family $({\bar A}_w)_{w\in\W}$
by
\be
{\bar A}_w\ :=\ (U({\bar A})_{p_1},...,U({\bar A})_{p_n})\in G^n=\A_w
\ee  
where $w=\{p_1,...,p_n\}$\footnote{It is here where we need Theorem 2.1.
When $U$ varies, the points ${\bar A}_w$ fill all of $G^n$.
Due to $G^n=\A_w$, every $U$ defines a point in $\A_w$.}.
It is not hard to see, that the result solves the consistency 
conditions and that every solution can be obtained in that way. 

The algebra $\Cyl$ contains a sub-algebra $\Cyl(\A/\G)$
of gauge invariant cylindrical functions. Also the results of 
\cite{MM,AL1,AL2,AL3} apply, with the graphs replaced by webs.
In particular, the spectrum $\overline{\A/\G}$ is 
\be
\overline{\A/\G}\ =\ \Ab/\Gb,
\ee
where $\Gb$ is the group of {\it all} the maps from $\S$ into $G$,
the Gel'fand completion of the gauge group acting in $\Ab$.
  
\subsection{A natural measure}
 A measure $\mu_0$ on the spectrum $\Ab$ is defined by any family of measures
$(\A_w,\mu_w)_{w\in\W}$ such that given a cylindrical
function $\Phi$ in the form (\ref{cyl}), the integral
\be
\int \phi d\mu_w\ =:\ \int_{\Ab}d\mu_0 \Phi
\ee
is independent of the choice of the web $w$. By the same arguments as
those used in \cite{AL1}, if $G$ is compact and we choose
for each $\mu_w$ the probability Haar measure, the resulting
family of measures does satisfy the consistency condition
and defines a natural, diffeomorphism invariant measure
on $\Ab$. The resulting integral restricted to the sub-algebra
of gauge invariant elements of $\Cyl$  coincides with
the natural measure defined in \cite{AL2,AL3}. This result 
was derived earlier by Baez and Sawin and our Theorem 2.1 mainly simplifies
the argument.   

\section{Spin-webs and diffeomorphism averaging}
\label{s3}

With the natural measure, the space of the cylindrical functions
is naturally completed to become the Hilbert space
\be
\H\ :=\ L^2(\Ab,d\mu_0).
\ee 
The aim of this section is to find a generalization of the 
spin-network states \cite{RS1,BS2} constructed from graphs. The
spin-networks  
gave an orthogonal decomposition of the Hilbert completion of
the subspace of $\H$ given by the piecewise analytic paths
and were used to define the averaging with respect to all
the diffeomorphisms of $\Sigma$.

\subsection{Spin-webs} 
\label{s3.1}

Given a web $w$ we  associate a Hilbert 
subspace $\H_w$, the Hilbert completion of $\Cyl_w$. 
That subspace is isometric with $L^2(G^n)$. That identification
provides an orthogonal decomposition given by $\L^2(G)=\oplus_jV_j$ 
where $V_j$ are  subspaces labeled by the irreducible representations
of $G$. In this way, to each {\it labelled web} $(w,j)$, 
that is, a pair consisting of a web $w$ and a labelling $j$ 
of the edges of $w$ by irreducible representations, we associate 
a finite dimensional subspace $V_{w,j}\subset\H_w$ called 
a {\it spin-web space}
\footnote{With an  edge $p$ of a web $w$ and an irreducible 
representation $j(p)$ one associates the linear span $V_{p,j(p)}$
of the the functions $f(A)={{\cal D}^{(j(p))}}^M_N(U(A))$, where
${{\cal D}^{(j)}}^M_N(U)$ stands for an $(M,N)$ entry of 
the matrix corresponding to an element $U\in G$ in the $jth$ 
representation and with respect to some fixed basis in the 
fundamental representation space. 
Then, $V_{w,j}=\oplus_{p}V_{p,j(p)}$, $p$ running through
the edges of $w$.}. The spin-web subspaces $H_{wj}$ 
of $\H_w$ corresponding to all possible labellings $j$
span $\H_w$, are finite dimensional and for different 
labellings they are orthogonal. 
 
If $j(p)=0$ for an edge $p$ of a labelled web $(w,j)$,
then we can consider another labelled spin-web $(w',j')$ given
by removing the edge $p$ from $w$ and maintaining the labelling
of the remaining edges. Obviously, the corresponding 
spin-web spaces are equal. In the case of graphs, 
that degeneracy can be removed by admitting only nontrivial
representations for the labelling of its edges and vertices.
However, in the case of the proper webs, that is not enough
and still there are spin-web spaces associated to
different webs $R(w')\not=R(w)$
such that $V_{wj}$ fails to be orthogonal to $V_{w'j'}$.
Since that is an important difficulty  in diffeomorphism averaging
of the spin-webs, we illustrate it by the following example.\\
{\bf Example.}  
\begin{figure}
\epsfxsize=15cm
\centerline{\epsfbox{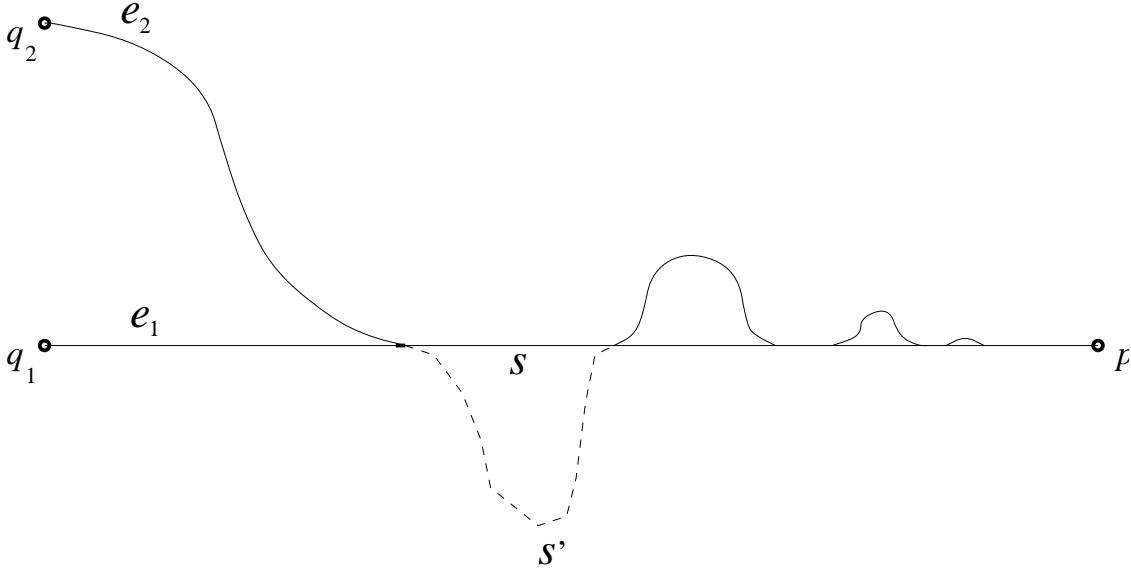}}
\vspace{2mm}
\caption{The edge $e_1$ is the horizontal line whereas the edge $e_2$
consists of the infinitely many bumps up the horizontal line connected with
the horizontal segments (only four bumps of $e_2$ are visible in the figure.)
The edges $e'_1$ and $e'_2$ are obtained by replacing the solid
segment $s$ by the dashed segment $s'$.}  
\end{figure}
The first web is $w=\{e_1,e_2\}$. This is just one tassel
of a base point $p$.  The second web consists of the deformed
edges. It  is  $w=\{e'_1,e'_2\}$. 
(see Fig.1). 
For the labellings of $w$ and $w'$
respectively we take $j(e_1)=j(e_2)=j'(e'_1)=j'(e'_2)={1\over 2}$.
Our claim is that the corresponding spin-web spaces $V_{w,j}$ and 
$V_{w',j'}$ are not orthogonal. To see this, consider a web $\tilde{w}$ 
obtained from $w$ and $w'$ by subdividing their edges,  such that the 
segments $s$  and $s'$ become path products of entire edges of 
$\tilde{w}$ (see Fig.2).           
\begin{figure}
  \begin{center}
    \epsfig{file=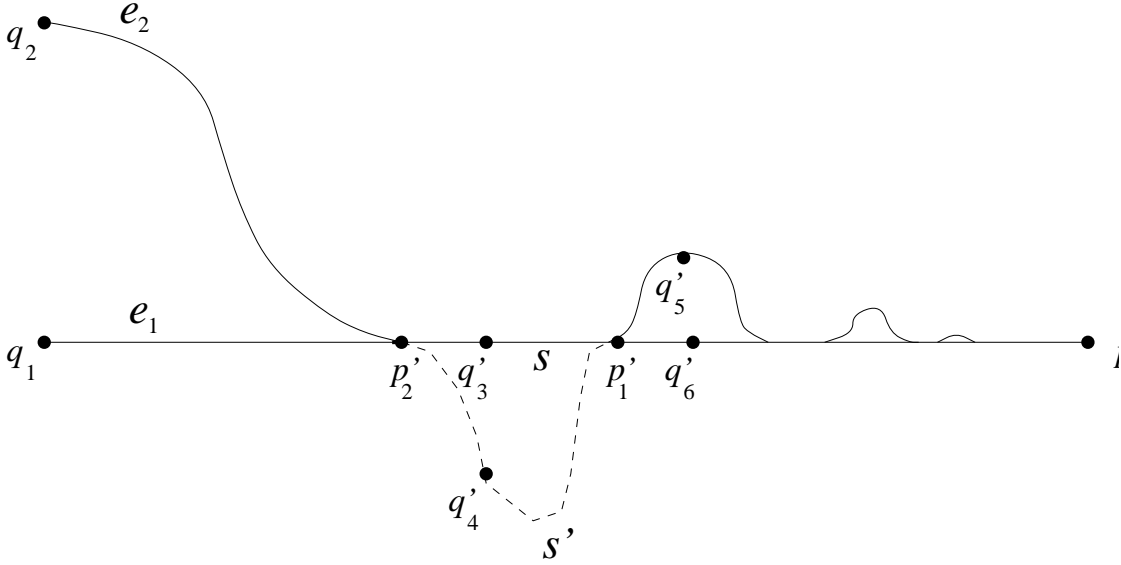,width=15cm}
  \end{center}
\vspace{2mm}
\caption{The web $\tilde{w}$ consists of three tassels. The first 
tassel is based at the point $p$ and consists of the segments connecting
$p$ with  the points $q'_5$ and $q'_6$. The second tassel is based at 
the point $p'_1$. It consists of the segments connecting  $p'_1$ with 
the points $q'_3, q'_4, q'_5, q'_6$. The third tassel is the set
of the segments connecting the point $p'_2$ with the points
$q_1, q_2, q'_3, q'_4$.}       
\end{figure}
The spin-web space $V_{w,j}$ is not any longer a spin-web space
with respect to the web $\tilde{w}$. It admits the non-vanishing
projection on the spin-web space given by the labelling $\tilde{j}$  
of $\tilde{w}$ which takes the value $0$ on the segments
of $s$ and $s'$, and ${1\over 2}$ otherwise. The same is true
for the spin-web $V_{w',j'}$.   
\medskip

The lesson the above example teaches us is that a smooth
deformation of a segment $s$ in $(w,j)$ such that it does not overlap the
remaining part of $w$ any longer does not necessarily result in a 
labelled spin-web $(w',j')$ such that
$V_{w',j'}$ and $V_{w,j}$ are orthogonal to each other. This would not be
true if we were working in the analytical category and with spin-networks
\cite{RS1,B2}. We could also see how two different webs define 
a refined, bigger one. 
  
Our goal now is to show that there exist certain ``non-degenerate" 
spaces $V_{w,j}$ with the property that they are orthogonal to
every other space $V_{w',j'}$ whenever the ranges of the webs do not
coincide, whether $(w',j')$ is also non-degenerate or not.
\begin{Definition} \label{def1}
Consider a web $w$ and a labelling $j$ of its edges with irreducible and 
nontrivial representations. Consider a regular point $x$ 
(see Sec. {s1.2}) of $w$ and  let the type of $x$ be such that the
edges $e_{I_1}, ..., e_{I_k}\in w$ are 
overlapping at $x$ . We will say that
$x$ is a {\bf non-degenerate point} 
whenever the trivial representation does  not emerge in the
decomposition of the tensor product of the representations 
$j(e_{I_1}), ..., j(e_{I_k})$.  Given a labelled web $(w,j)$, if 
every regular point is nondegenerate, then we will say that 
$(w,j)$ is {\bf nondegenerate}. 
\end{Definition} 
We then have the following result.
\begin{Theorem} \label{th2}
Suppose $V_{w,j}$ is the spin-web space assigned to
a non-degenerate labelled web $(w,j)$ and  the spin-web space $V_{w',j'}$ 
assigned to $(w',j')$ is not orthogonal to  $V_{w,j}$; 
assume also that $j'(e')\not=0$ for every $e'\in w'$; then:

$(i)$ The labelled spin-web $(w',j')$ is also non-degenerate; 

$(ii)$ The ranges of the spin-webs coincide,
\be
R(w) = R(w');
\ee

$(iii)$ There is a web $w''\ge w,w'$ which can be obtained
from each of the webs $w,w'$ by subdividing its edges.  

\end{Theorem}
Proof of Theorem \ref{th2} :\\
The key fact is the existence of a web $w''$ such that 
\be
w''\ge w,w'.  
\ee   
One has to remember (see Example of Sec. \ref{s3.1} ), however, that for 
proper webs, that is for webs which are not just graphs, given the 
spin-web space $V_{w,j}$
and the bigger web $w''$,  the space  $V_{w,j}$ is
not in general one of the spaces $V_{w'',j''}$ with some
labelling $j''$. All that we know  is that in general
\be
V_{w,j}, V_{w',j'}\ \subset\ \oplus_{j''}V_{w'',j''},
\ee 
where $j''$ is some finite set of labellings.
Since the spin-web spaces $V_{w,j}$ and $V_{w',j'}$ are not orthogonal
to each other, there is a labelling $j''_0$ of the web $w''$ such that
both the projections 
\be\label{w''}
V_{w,j}\ \rightarrow\ V_{w'',j''_0},\ \ \ and \ \ \  
V_{w',j'}\ \rightarrow\ V_{w'',j''_0}
\ee
are not trivial.  An easy 
observation is the following: 
\begin{Lemma} \label{la1}
Let $(w,j)$ and  $(w',j')$ be as in the hypothesis above;
there is a web $w''_0\ge w,w'$ which can be obtained by sub-dividing
the edges of $w'$.  
\end{Lemma}
Proof of Lemma \ref{la1} :\\ 
Let $w''$ be a web such that $w''\ge w, w'$. Let $w_1, w_2 \subset w''$
be webs which consist of the segments of the edges of $w$, and 
respectively, $w'$. Thus we can write,
\ba
w_1\ =\ \{e''_1,...,e''_k, a''_1,...,a''_n\},\nonumber\\
w_2\ =\ \{e''_1,...,e''_k, b''_1,...,b''_m\},\nonumber\\
w''\ =\ \{e''_1,...,e''_k, a''_1,...,b''_m, c''_1,...\}, 
\ea  
that is the webs $w_1$ and $w_2$ share the edges $e''_I$,
the edges $a''_I$ denote those of the edges of $w_1$
which are not contained in $w_2$, whereas $b''_J$
denote the edges of $w_2$  not contained in $w_1$.
The first observation is, that the  web obtained from the
third web above by removing the extra edges $c_I$, that is,
\be
w''_0\ :=\  \{e''_1,...,e''_k, a''_1,...,b''_m\}
\ee
also satisfies $w''_0\ge w,w'$. It is constructed
from segments of the edges of $w$ and $w'$. 
Consider now the labelling $j''_0$ of (\ref{w''}). 
Since the first projection is not trivial and due to
the non-degeneracy of $(w,j)$ we have
\be\label{j} 
j''_0(e''_I), j''_0(a''_J)\ \not=\ 0, \ \ \ and \ \ 
j''_0(b''_I) =\ 0.
\ee 
On the other hand, because the second projection is  
non-trivial either, we have
\be
j''_0(a''_J)\ =\ 0.
\ee
Therefore, the extra elements $a''_I$ of the web $w_1$
do not exist, and we finally have
\ba\label{ww}
w_1\ =\ \{e''_1,...,e''_k\},\nonumber\\
w''_0\ =\ w_2\ =\ \{e''_1,...,e''_k, b''_1,...,b''_m\}.\\
\ea 
This completes Lemma \ref{la1}.\\
$\Box$\\
\\
>From now on, let 
\be
w''=w''_0\ge w,w'
\ee 
be the web given by Lemma \ref{la1}. 
Eventually, we will show that if  spin-web spaces $V_{w,j}$ and $V_{w',j'}$
are not orthogonal to each other, then the non-degeneracy of $(w,j)$
implies the non-degeneracy of  $(w',j')$. Together with Lemma \ref{la1} 
that will complete the proof. But before that we need two intermediate 
lemmas. 

Consider two webs $w\le w''$.  An  edge $e\in w$ is the path 
product of the elements of $w''$,
\be
e\ =\ {e''}_{k}\circ ... {e''}_{1},
\ee
where ${e''}_{1}\in w''$ is the one that contains the base point of $e$.
We will see below that given a labelling $j$ of the web $w$, 
in the decomposition of the corresponding spin-web space $V_{w,j}$
into the spaces $V_{w'',j''}$ the labelling $j''$ of the beginning
segment ${e''}_{1}$ coincides with the labelling $j$ of the corresponding
edge $e$, for every edge $e\in w$.  
\begin{Lemma} \label{la2}
Let  $(w,j)$ be a labelled web and $w''\ge w$; consider an edge 
$e\in w$; let ${e''}_{1}\in w''$ be the segment of $e$ containing its 
base point; then
\be
V_{w,j}\ \subset\ \oplus_{j'':j''({e''}_{1})=j(e)}V_{w'',j''}.
\ee
\end{Lemma}
Proof of Lemma \ref{la2} :\\ 
The spin-web space $V_{w,j}$ is spanned by
the cylindrical functions of the form
\be
\otimes_{e\in w} {{\D}^{(j(e))}}_{n(e)}^{m(e)}(U_{e}). 
\ee
Each such function is expressed by the parallel transports 
along the elements of $w''$ through the decomposition:
\be\label{prod}
{{\D}^{(j(e))}}_{n(e)}^{m(e)}(U_(e))\ =\ \
({\D}^{(j(e''_{k}))}(U_{e''_{k}})...
{\D}^{(j(e''_{1}))})(U_{e''_{1}}))_{n(e)}^{m(e)}
\ee
where $j(e''_{1})=...=j(e''_{{k}''})=j(e)$ since 
${\cal D}$ is a representation and $U_{e''_{k}}...U_{e''_{1}}=U(e)$.  

Notice that any of the factors 
${\D}^{(j(e''_{1})})_{n(e)}^{m(e)}(U_{e''_{1}})$
appears only once in the whole product (\ref{prod}). To see this, we 
note that different edges $e_1$ and $e_2$ may share the same segment 
$e''_I$. However, by the definition of a web $w$ (\cite{BS}, 
see also section \ref{s1.2}), a segment of an
edge $e_1$ of $w$ connected to the base point of a tassel in $w$ cannot be 
contained in a different 
edge $e_2$ of the same web.\\
$\Box$  
\begin{Lemma} \label{la3} Suppose $(w,j)$ is a nondegenerate labelled web
and $(w',j')$ is another labelled web such that
the associated spin-web spaces $V_{w,j}$ and $V_{w',j'}$
are not orthogonal; then  $(w',j')$ is also nondegenerate.
\end{Lemma}
Proof of Lemma \ref{la3} :\\
Consider a regular point $x'$ of the web $w'$. Denote by 
$e'_{1},..., e'_{k}$ the edges of $w'$ which overlap at $x'$.  
What  we have to show is that the tensor product of the corresponding 
representations $j'(e'_{1})$, ... , $j'(e'_{l})$ does 
not contain  the trivial representation. 
Use  again the web $w''$ of Lemma \ref{la1} and consider the labelling 
$j''_0$ of (\ref{w''}). Let the sub-webs $w_1, w_2\subset w''$
be defined as in (ref{ww}).
One of the edges of $w''$ is the segment of the edge $e'_1$ which
contains its base point of the corresponding tassel.   
Denote it by $e''$. Notice, that according to Lemma \ref{la2}
\be
j''_0(e'')\ =\ j'(e'_1)\ \not=\ 0.
\ee
Comparison with (\ref{j}) shows, that $e''$ is one of the edges
$e''_I$, say $e''_1$, of the sub-web obtained by subdividing the 
edges of $w'$.  The same argument applies to each of the 
edges overlapping at $x'$. Denote the corresponding segments by
$e''_1,...,e''_l\in w_1\subset w''$. Since they are segments 
of edges $e'_I$ of a single tassel and each of them contains the base
point, there is a regular  point $x''$ of $w''$ such the edges
$e''_1,...,e''_l$ overlap at $x''$ and $x''$ is not intersected
by any other of the edges of $w''$. But the tensor product of the 
representation  $j''_0(e''_1),...,j''_0(e''_l)$ does not 
give the trivial one, because the labels $j''(e_I)$ come from the
decomposition of the non-degenerate labelled web $(w,j)$
into the spin web spaces $V_{w'',j''}$ and the projection
onto $V_{w'', j''_0}$ is non-trivial. This concludes the
proof of Lemma \ref{la3} and the proof of Theorem \ref{th2}. \\
$\Box$\\
\\
What we learnt from this section is that despite of the difficulties
shown in Example of Sec. \ref{s3.1} there is a class of spin-webs
which have the property, that a space deformation of 
a given spin-work results, generically,  in an orthogonal spin-web space.

\subsection{Averaging over the diffeomorphism group}
\label{s3.2}
 
In \cite{ALMMT2} the spin-networks of the analytic category can be employed 
to construct solutions to the diffeomorphism constraint\footnote{
That is, diffeomorphism invariant state which are elements of the space
$\Cyl'$, the topological dual of $\Cyl$.}. 
The construction consists
of averaging the state over the diffeomorphisms of $\Sigma$.
With the Baez-Sawin results \cite{BS} the averaging was easily 
extended to  averaging of a spin-network state with respect
to the smooth diffeomorphisms \cite{BS2,LM}.    
Our goal now is to extend that construction to  a spin-web state. 

We need first some notation.

Given a labelled web $(w,j)$ the gauge group acts in $V_{w,j}$
at the end points $x,y$ of an edge $p$ through its action
on the parallel transport along the edges,
\be
U_e\ \mapsto\ g(y)^{-1} U_p g(x)
\ee
where the $G$-valued function $g:\Sigma\rightarrow G$ defines
an element of the group of the gauge transformations. 
The end points of the edges of a web will be called vertices 
in the sequel.

One can further decompose
each $V_{w,j}$ into the irreducible representation $V_{w,j,l,c}$,
of that action, assigning to each vertex $v$ of the web $w$
an irreducible representation $l$ and  an extra label $c$,
because several equivalent but mutually orthogonal representations can appear 
more then once (see \cite{RS1,B2,ALMMT2}  for the details about
the gauge invariant spin-networks and  \cite{AL4,Thvol,Thloop}). The quadruple
$(w,j,l,c)$ will be also called a {\it labelled} web.
                                        
As in \cite{ALMMT2} an important notion will be the group of trivial 
action diffeomorphisms of a web whose vertices are labelled
by irreducible representations. We define it as follows. 
Given a web $w_0$ fix a labelling $l_0$ of its vertices
by the irreducible representations. The group $\TA(w_0,l_0)$ is the 
subgroup of those  diffeomorphisms of  $\Sigma$ which act trivially in 
the space $V_{w_0,j,c,l_0}$ for every labelled web  $(w_0,j,l_0,c)$
with the fixed web $w_0$ and the labelling $l_0$,
and with arbitrary labellings $j,c$. In fact, the group depends quite
weakly on $l_0$.  If  the representation $l_0(v)$ is
non-trivial for every  vertex $v$, then the group 
$\TA(w_0,l_0)$ consists of those diffeomorphisms which preserve every 
edge (including its orientation) of $w_0$.  However, there are some 
more diffeomorphisms acting `trivially' from the above point of view.   
Consider a 2-valent vertex $v_0$ 
of the web $w_0$ which is the intersection  point of two edges  $e_1,e_2\in 
w$   such that, modulo change of orientation, one edge is a smooth  
extension 
of the other. Suppose $l_0(v_0)$ is the trivial representation. Then,
every element of the associated space $V_{w_0,j,c,l_0}$, as a function
on $\A$, depends on the connection $A$ through the parallel transport
$U_{e_2\circ {e_1}^{-1}}$.  Therefore in this case, every diffeomorphism 
preserving the oriented non-parameterized curve $e_2\circ {e_1}^{-1}$ is an 
element of $\TA(w_0,l_0)$.

The group of smooth diffeomorphisms of $\Sigma$ acts naturally
in the space of the cylindrical functions $\Cyl$. Given
a diffeomorphism $\varphi$ and $\Psi\in\Cyl$  denote 
the result of the action by $\varphi \Psi$.  Consider  a labelled web  
$(w,j,l,c)$, and $\Psi\in V_{w,j,l,c}$. We attempt to define an averaged 
state $\langle \Psi|_{\av}$ by the following action on a cylindrical 
function $f\in \Cyl$:
\be\label{av}
\langle\Psi|(f)\ :=\ {1\over \kappa(w,j,l)}
\sum_{[\varphi]\in\Diff(\Sigma)/\TA(w,l)}
(\varphi\Psi|f),
\ee          
where $\kappa(w,j,l)$ is a  constant. The following two conditions
should be satisfied by the averaging:

$(i)$ finiteness; given $\Psi$  the result should be finite for a large 
class of $f\in \Cyl$;

$(ii)$ consistency; since the spin-web spaces are not orthogonal
to each other, we have to ensure that given two different
labelled webs $(w,j)$ and $(w',j')$ and a state $\Psi\in V_{w,j}
\cap V_{w',j'}$ our definition of the averaging gives
the same result regardless of which of the above webs we use.

Suppose $f$ is contained in the subspace $V_{w',j'}$ associated with
some labelled web $(w',j')$ and that $j'(e')\not=0$ for every
$e'\in w'$.  Now, owing to theorem \ref{th2},
the only nonzero terms in (\ref{av}) come from the diffeomorphisms
such that both, $\varphi(R(w))=R(w)$ and  the webs $w$ and $w'=\varphi(w)$
satisfy the property $(iii)$ of Theorem \ref{th2}. Define the 
{\it symmetry group} $\SG(w,l)$ of a labelled web $(w,j,l,c)$ to be
\ba
\SG(w,l)\ = \{[\varphi]\in \Diff(\Sigma)/\TA(w,l)| 
\varphi(R(w))=R(w)\nonumber\\
 {\rm and}\ \  w':=\varphi(w)\ \ {\rm satisfies}
\ \ (iii)\ \ {\rm of\ \  Theorem \ref{th2}}\}.\nonumber\\
\ea     
What we need to assume is that $\SG(w,l)$ is finite. A 
general cylindrical function can be always written as an infinite sum,
\be
f\ =\ \sum_{j'} a_{j'} f_{j'},\ \ f_{j'}\in V_{w',j'},
\ee
with respect to the labellings $j'$ of the edges of a single web $w'$. 
To see that also for a general cylindrical function $f$ the sum 
(\ref{av}) is finite as long as the symmetry group $\SG(w,l)$ is finite, 
consider  $[\varphi]$ in the sum (\ref{av}) such that
\be
(\varphi\Psi|f) \ \not=\ 0. 
\ee    
The projection of $\varphi\Psi$ onto $V_{w',j'}$
is not trivial only for finite number of the labellings $j'$.
(To see that, take $w'\ge w$, to represent $f$). 
 
Turn now to the consistency condition $(ii)$. It will constrain
the freedom of the constants $\kappa(w,j,l)$. Consider a labelled
web $(w_0,j_0,l_0,c_0)$ and a state
\be
|w_0,j_0,l_0,c_0,M_0>\ \in\ V_{w_0,j_0,l_0,c_0}
\ee
 which satisfies an identity
\be\label{id2}
a_0|w_0,j_0,l_0,c_0,M_0>\ =\ a_1 |w_1,j_1,l_1,c_1,M_1>\ +\ ...\ +\ 
 a_r|w_r,j_r,l_r,c_r,M_r>.
\ee    
Without loss of generality,
we assume that the identity cannot be obtained as a sum of two non-trivial 
identities.  That is we may assume that $|w,j,l,c,M>$ on the left hand
side is not orthogonal to any of the terms on the right hand side.
Therefore, the web $w$ is in the relation $(iii)$ of Theorem \ref{th2}
with each of the webs $w_1, ..., w_r$ on the right hand side. 
Moreover, if $l_s(v_s) \not= 0$ for any of the vertices $v_s$
of any of the webs above, then the point $v_s$ is a vertex of
every of the webs in the identity, and 
\be
l_1(v_s)\ =\ ...\ =\ l_r(v_s).
\ee        
We may also write the identity in such a way, that the labelled
webs $(w_0,j_0), ..., (w_r,j_r)$ are all different.

 To ensure that the  averaging is consistent with the identity
we observe the following:

\begin{Lemma}\label{lb1} Suppose a labelled spin-web $(w,j,l,c)$
is non-degenerate. If $(w',j',l',c')$ is another labelled spin-web
such that the corresponding spin-web spaces $V_{w,j,l,c}$ and
$V_{w',j',l',c'}$ are not orthogonal to each other,
then the corresponding  groups of the trivially acting diffeomorphisms 
coincide,
\be
\TA(w,l)\ =\ \TA(w',l').
\ee   
\end{Lemma}
Proof of Lemma \ref{lb1} :\\
We again take advantage of Teorem \ref{th2}. Let $w''$ be the web
given by Theorem \ref{th2} which satisfies $(iii)$ of that theorem.
Consider the following labelling $l''$ of the vertices of 
$w''$ with the irreducible representations. For a vertex
$v$ of $w''$ set
\be
l''(v)\ =\ \cases{l(v),& if $v$ is a vertex of $w$,\cr
                   0, & otherwise.\cr}
\ee
Recall that every vertex $v$ of $w''$ which is not a vertex 
of $w$ is obtained by subdividing an edge of $w$. 
It is easy to check, that the corresponding 
groups of trivially acting diffeomorphisms coincide,
\be
\TA(w,l)\ =\ \TA(w'',l'').
\ee 
But  it is also true for $\TA(w',l')$ and  $\TA(w'',l'')$,
which completes the proof of Lemma \ref{lb1}. \\
$\Box$\\
\\
The formula (\ref{av}) for the averaging depends 
on the group $\TA(w,l)$ of a given labelled web $(w,j,l,c)$
and on the unspecified constant $\kappa(w,j,l)$. 
For every term in the identity $(\ref{id2})$ the group 
of the trivially acting diffeomorphisms is the same, and equals 
$\TA(w_0,l_0)$. Therefore, to ensure the consistency, we have
to take the constants $\kappa(w,j,l)$ depending only
on $\TA(w,l)$,
\be\label{av2}
\kappa(w,j,l)\ =\ f(\TA(w,l)).
\ee
A quite natural choice is for example
\be\label{av3}
\kappa(w,j,l)\ =\ |\SG(w,l)|,
\ee 
where $|\SG(w,l)|$ is the number of elements of the symmetry
group of a labelled web.   

Let us summarize our results:
Denote by $\T_{{\rm nondeg}}$ and  $\T_{{\rm deg}}$ the subspaces
of $\Cyl$ spanned by the  spin-web subspaces corresponding to,
respectively, non-degenerate and degenerate labelled spin-webs.
Next, let  $\T_{nondegfin}$ and  $\T_{nondeginf}$ be spanned by the spin-web
spaces corresponding to nondegenerate labelled webs 
of, respectively, finite and infinite symmetry groups. 
\begin{Theorem}\label{th3}
The Hilbert space $\H$ of all the square integrable functions,
can be orthogonally decomposed in the following way,
\be\label{decomp}
\H\ =\  \overline{\T_{{\rm nondeg}}}\oplus  \overline{\T_{{\rm deg}}}\ =\ 
 \overline{\T_{nondegfin}}\oplus  \overline{\T_{nondeginf}}\oplus  
 \overline{\T_{{\rm deg}}}, 
\ee
where the over lining denotes the Hilbert closure;    
the diffeomorphism averaging given by (\ref{av},\ref{av2}) defines 
a linear map
\be
\T_{nondegfin}\ni\Psi\ \mapsto\ \langle\Psi|\in\Cyl', 
\ee 
$\Cyl'$ being the topological dual to the space $\Cyl$. 
\end{Theorem} 

The proof follows from the above arguments, Theorem \ref{th2}
and Lemma \ref{la3}.  
   
Some remarks are in order about the subspaces $\T_{nondeginf}$ and  
$\T_{{\rm deg}}$ which are excluded from the domain of the averaging
map.

The first subspace  is given by non-degenerate 
spin-webs with infinite symmetry groups. An example of such a spin-web
was given in \cite{BS2}. The spin-web considered there is degenerate in
terms of the present paper, but it is not hard to modify it to
a nondegenerate spin-web of the infinite symmetry group. 
The idea of the construction of such an example is that a part of a web 
can be non trivially mapped into
itself by a diffeomorphism which preserves the rest of the web.
If similar parts emerge infinitely many times (in every neighborhood
of a base of a tassel) then we have infinitely many diffeomorphisms.
which are symmetries of the web. 

The second group of states is given by degenerate labelled 
webs $(w,j)$. An example of such a web is shown in Fig. 1.
In all the examples of degenerate labelled webs we are aware of, 
an associated cylindrical function can be decomposed into a converging  
infinite sum of cylindrical functions each being associated to a web  
obtained from the original one by removing a (degenerate) segment.  
See section \ref{s4.3}.

>From the point of view of a diffeomorphism invariant theory of connections, 
the averaged states are solutions of the diffeomorphism constraint.   
The action of the Hamiltonian constraint operator of the gravitational field 
can be extended to those states. We go back to that issue in Sec. \ref{s5}.   

\section{Diffeomorphism invariant operators}
\label{s4}

In the analytic category \cite{ALMMT2}, the Hilbert space 
of cylindrical
functions is orthogonally decomposed into subspaces labeled
by graphs in such a way, that every diffeomorphism invariant 
operator preserves the subspaces.                                 
The aim of this subsection is to show, that in the smooth case,
certain subspaces are also necessarily preserved by 
diffeomorphism invariant operators. However, the result that
we derive below is not as strong as in the analytic category.

\subsection{Ranges of diffeomorphism-invariant operators}
\label{s4.1}

Given a web $w$   denote by $\H_{R(w)}$  the Hilbert spaces 
obtained by using all the cylindrical functions 
which are associated to the webs  whose ranges are contained in
the range of $w$,
\be
\H_{R(w)}\ :=\ \overline{\oplus_{w'|R(w')\subset R(w)} \H_{w'}}. 
\ee 
\begin{Theorem} \label{th4}
Suppose  $\O$ is a diffeomorphism invariant
operator defined in a domain $\D(\O)$ in the Hilbert space
$\H\ =\ L^2(\A,d\mu_0)$; then, 

$(a)$ for every web $w$, the Hilbert subspace $\H_{R(w)}$ associated 
to its range is preserved by the operator $\O$, 

$(b)$ in particular, when a web is a graph, then the space 
\be
\overline{\oplus_{j}V_{w,j}},
\ee
$j$ running over over all the labellings of the edges of $w$
by the irreducible (possibly trivial) representations, 
is preserved by the operator $\O$,  

$(c)$ suppose, that for every labelled web $(w',j')$, it is true that,
$V_{w',j'}\ \subset\ \D(\O^\dagger)$; let $\Psi\in V_{w,j}\cap\D(\O)$
and $\N\in V_{w',j'}$ where $(w,j)$ are labelled webs;
then,  
\be
(\O\Psi|\N)\ \not=\ 0,\ \ \ \Rightarrow\ R(w')=R(w),
\ee
provided the labellings $j,j'$ do not assign the trivial 
representation to any of the edges of the webs $w$ and, 
respectively, $w'$; in particular, if the web $w$ is a graph,
then the operator preserves the space
\be
\oplus_{j}V_{w,j},
\ee
 $j$ running through all the labellings of the edges of $w$
by irreducible and non-trivial representations. 
\end{Theorem}

Proof :\\               
Consider a web $w$ and a state $\Psi\in \H_{R(w)}\cap \D(\O)$.
For the proof of $(a)$, it is enough to show, that for every 
labelled web $(w',j')$ such that $w'\ge w$ containing an edge $e'\in w'$ 
such that the range $R(e')$ is not contained in $R(w)$ 
\be  
j'(e')\ \not= 0\ \Rightarrow\ \O\Psi\perp V_{w',j'}.
\ee    
(Notice, that the subspaces $V_{w',j'}$ span a dense subspace in the 
whole of $\H$, even when we restrict ourselves, to $w'\ge$ a given $w$ 
because we allow $j'$ to contain trivial representations.)  
Let $e'_0\in w'$ be an edge such that the range $R(e'_0)$ is not contained 
in $R(w)$. In $e'_0$, there is a regular point $x'_0$ of the web $w'$ which 
is not contained in $R(w)$. According to the definition
of a tassel, the edge $e'_0$ contains a sequence of the regular
points $(x'_k)_{k=0}^{\infty}$ such that each of the points
has the same type as $x'_0$, and the sequence converges to the 
beginning point of $e'_0$, the base point of the corresponding tassel.  
For each of the points $x'_k$ take an open neighborhood 
$\U_k\ni x'_k$ diffeomorphic to a ball, such that the intersection 
$R(w')\cap \U_k$ is an embedded open interval, say $s'_k$. 
Consider a diffeomorphism $\varphi\in \Diff$  such that 

$i)$ $\varphi$ preserves each of the   $\U_k$ and acts trivially
outside of $\cup_k\U_k$,

$ii)$ $\varphi(x'_k)\ \not=\ x'_k$, for every point of the
sequence,

$iii)$ $\varphi$ deforms each segment $s'_k$ in the same way, modulo
diffeomorphisms; that is, for every pair of the segments 
$s'_k$ and $s'_l$, the pair $(s'_k,\varphi(s'_k))$ is diffeomorphic
to $(s'_l, \varphi(s'_l))$.  

Let $e'_1,...,e'_m$ be the edges of $w'$ which overlap
the edge $e'_0$ at $x_0$. Hence, they overlap it at each of the 
points $x_1,...,x_k,...$.
It follows from ($i-iii$) above, that the family of edges
obtained from $w'$ and its deformation, namely
\be
w''\ :=\ w'\cup \varphi(w')\ =\ \{\varphi(e'_0),\varphi(e'_1),...,
\varphi(e'_m)\}\cup w',
\ee
is also a web. 

Now, given a labelling $j'$ of $w'$, we can trivially extend it to
a labelling $j''$ of $w''$ by setting
\be
j''(\varphi(e_0'))\ =\ ...\ j''(\varphi(e_m'))=\ 0. 
\ee
On the other hand, the diffeomorphism $\varphi$ carries $j'$ into 
another labelling $\varphi(j')$ of the web $w''$ given by
$\varphi(j')(\varphi(e'):=j'(e')$ for every edge of $w'$. Obviously 
we then have   
\be\label{nonzero}
j'(e'_0)\ \not=\ 0\ \Rightarrow\ \varphi(j')\ \not=\ j''.
\ee
Therefore, if $j'$ satisfies (\ref{nonzero}) then for every  
$\N\in V_{w',j'}$, the functions
$\N$ and $\varphi(\N)$ are orthogonal to each other.
But, the diffeomorphism acts trivially on $\Psi$.
Therefore from the diffeomorphism invariance of $\O$,
the projection of $\O\Psi$ on $\N$ is $\varphi$-invariant,
\be
(\O\Psi | \varphi \N) \ =\ (\varphi^{-1} \O\Psi | \N)
\ =\  (\O\varphi^{-1} \Psi | \N) \ =\ (\O \Psi | \N).
\ee 
 Moreover, it is easy to iterate the above construction
to obtain an infinite family of diffeomorphisms each of which 
satisfies the conditions $(i-iii)$ above and    
such that for every pair of this set of diffeomorphisms,
\be
\varphi' \not=\ \varphi\mbox{ it follows that }
(\varphi'\N_{w'',p''} | \varphi \N_{w'',p''})\ =\ 0.
\ee
If $\varphi$ varies through all possible diffeomorphisms like this,
the set of the resulting states $\varphi\N$ has an infinite number
of mutually orthogonal elements. Since a densely defined operator cannot 
have an infinite
number of non-vanishing matrix elements of the same value it follows that 
\be
(\O\Psi | \varphi \N_{w'',j''})\ =\ 0.
\ee 
This is sufficient to complete the proof of part $(a)$ of 
the Statement.\\

If $w$ is a graph, then $R(w')\subset R(w)$ implies that 
$w'$ is also a graph. Let $v'$ be a vertex of $w'$ which
is not a vertex of the graph $w$. Therefore, it is contained in
the interior of an edge $e\in w$.  For every labelling $(w',j',l',c')$
of the graph $w'$ such that $l'(v')\not= 0$, and for every 
spin-web state $\N\in V_{w',j',l',c'}$, we  prove that
\be
(\O \Psi| \N)\ =\ 0,
\ee
in the same way as before. We construct an infinite class of the 
states diffeomorphic to $\N$ and orthogonal to each other.   
The diffeomorphisms we use move the vertex $v'$ along the edge $e$
but preserve the edges of the web $w$, hence they also preserve 
the function $\Psi$. This completes the proof of the part $(b)$.
The point $(c)$ follows easily from $(a),(b)$ and their applications
to the adjoint operator $\O^\dagger$. This completes the proof
of Theorem \ref{th4}.\\       
$\Box$

\subsection{Strongly degenerate labelled webs}
\label{s4.2}

In the previous section we have spited the Hilbert space 
into the orthogonal sum of degenerate and nondegenerate
spin-web states and the last sector has again been orthogonally
decomposed with respect to possible symmetry groups of
the labelled spin-webs. The diffeomorphism averaging
is well defined only in one of those sectors, namely in
the subspace $\T_{nondegfin}$ spanned by all the spin-web states
of a non-degenerate labelled webs who have at most
finite symmetry groups. For this reason it is natural to
ask if a diffeomorphism invariant operator  is necessarily
consistent with that decomposition and whether it
preserves the domain of the diffeomorphism averaging.

It is easy to see, that owing to Theorem \ref{th4}.$c$,
a symmetric diffeomorphism invariant operator preserves 
the finiteness of the symmetry groups.  Regarding the non-degeneracy,
we will show below, that the image of a non-degenerate spin-web state 
under the action of a diffeomorphism invariant operator
is orthogonal to what we will define as `strongly degenerate' 
states. We will also indicate a large class of webs for which 
the degeneracy is equivalent to the `strong degeneracy', and we will 
conjecture, that the equivalence holds for all the spin-webs.  

Given an edge $e$ let        
\be
\hat{p}_e :\H\ \rightarrow\ \H
\ee
be the projection onto the subspace given by the cylindrical functions 
independent of the holonomy along $e$. Explicitely, the projection
can be defined as follows. Given a cylindrical function $\Psi\in \Cyl$
take a web $w\ge \{e\}$ such that $\Psi\in \H_w$. Let $e_1,...,e_k\in w$
be all those elements of $w$ which are the segments of $e$. We will use
the orthogonal projection,
\be
\hat{p}_{e,w} : \H_w \to \oplus_{j:j(e_1)=...=j(e_k)=0}V_{w,j},
\ee
to define
\be
\hat{p}_e \Psi\ =\ \hat{p}_{e,w}\Psi.
\ee 
 
A labelled web $(w,j)$ is {\it strongly degenerate} if 
there is an infinite sequence of disjoint regular segments 
$s_1,...,s_l,...$ such that for every $\Psi\in V_{w,j}$
\be\label{strdeg}
\lim_{l\rightarrow\infty}(1-\hat{p}_{s_1})...(1-\hat{p}_{s_l})
\Psi\ =\ 0,
\ee 
in the Hilbert space topology. 

Indeed, the strong degeneracy implies the week degeneracy.
A labelled web $(w,j)$ is either degenerate or non-degenerate.
If it is non-degenerate, then, for every regular segment $s$
\be
\hat{p}_s(V_{w,j})\ =\ 0,
\ee 
and the left hand side of (\ref{strdeg}) is $\Psi$ itself
for every $\Psi\in V_{w,j}$.  
  
\begin{Theorem}\label{th5} Suppose $\O$ is a diffeomorphism invariant
operator in $\H$ such that the domain $\D(\O^\dagger)$ contains all the 
spin-web subspaces $V_{w',j'}$. Then, for every non-degenerate spin-web 
state $\Psi$ and every strongly degenerate labelled spin-web state
$\N$,
\be
(\O\Psi|\N)\ =\ 0.
\ee    
\end{Theorem}
Proof :\\
It follows from the non-degeneracy of $\Psi$ and from Theorem
\ref{th2}, that any diffeomorphism $\varphi$ which deforms only the 
segment $s_1$ and acts trivially in $R(w)\setminus R(s_1)$ 
satisfies
\be
(\varphi \Psi | \O^\dagger \hat{p}_{s_1} \N)\ =
\ (\Psi|\O^\dagger\hat{p}_{s_1} \N),\ \ \
(\varphi\Psi|\Psi) =\ 0..
\ee  
So, there exists an infinite family of states $\varphi_k\Psi$, such that  
\ba
(\varphi_{k_1}\Psi | \O^\dagger\hat{p}_{s_1}\N) \ =\ 
(\varphi_{k_2}\Psi | \O^\dagger\hat{p}_{s_1}\N),\noindent\\
(\varphi_{k_1}\Psi|\varphi_{k_2}\Psi)\ =\ 0.
\ea 
It follows, that
\be
(\O\Psi|\hat{p}_{s_1}\N)\ =\ 0.
\ee
That shows, that
\be
(\O\Psi|\N)\ =\ (\O\Psi | (1-\hat{p}_{s_1})\N).
\ee
Next, using the same argument for the sequences $s_2,.,s_k,...$ we prove, 
that for arbitrary $k$
\be
(\O\Psi|\N)\ =\ (\O\Psi|(1-\hat{p}_{s_k})...(1-\hat{p}_{s_1})\N).
\ee
Taking limit $k\rightarrow 0$ and using the strong degeneracy of $\N$
completes the proof of Theorem \ref{th5}.\\
$\Box$\\
\\

Our conjecture is: 
\medskip

\noindent{\bf Conjecture.} {\it Every degenerate labelled web is
strongly degenerate.}
\medskip

So if the conjecture is true, then every  symmetric, diffeomorphism invariant 
operator   preserves the subspace $\H_{nondegfin}$
in which the diffeomorphism averaging has been well defined.

We will advocate the conjecture in the next subsection.

\subsection{A large class of strongly degenerate spin-webs.}
\label{s4.3}

There is a large class of webs, for which we can prove
that the degeneracy  of a labelling does imply the strong degeneracy. 
Consider throughout this section the  webs for which the following is true:
For every tassel $T=\{e_1,...,e_m\}$ of a given web $w$ and for 
a parameterization of the edges  of $T$  given by the  
point $(ii)$ of the definition of tassel   (say, the parameterization
is defined on $[0,1]$) for every $0<t<1$ there is $0<t'<t<1$ such that 
the segments $e_{1tt'}, ..., e_{mtt'}$ of all the edges $e_1,...,e_k$ 
are holonomically indepenent
(The above property is satisfied for the web in Example given
in Sec \ref{s3.1}. In fact, we do not know of any counter example, any  
web that would fail to satisfy that property.)
 
Now, consider a degenerate labeling  $(w,j)$ of a web $w$ and
a state $\Psi\in V_{w,j}$. We will calculate explicitely
the action of the projection operator $\hat{p}_s$ for appropriately
selected segments. Let $s$ be a regular segment, such that 
the tensor product $\rho_{j(e_1)}\otimes...\otimes \rho_{j(e_k)}$
contains the trivial representation in the decomposition into 
the irreducible representations, where $e_1,..., e_k$ are 
the edges of $w$ that overlap
$s$. The degeneracy guarantees the existence of such a segment. 
Then, owing to the properties of a tassel, there is  
a segment $s_0$ of the same type as $s$ which is close 
enough to the base point of the tassel for the following to be true:
If we write each of the edges overlapping at $s$ as
\be
e_I\ =\ e_{I+}\circ s\circ e_{I-}, \ \ \ I=1,...,k,
\ee    
then the segments $e_{1+},...,e_{k+}$ are holonomically independent
(the beginning segments $e_{I-}$ are  holonomically  independent for any 
choice of $s$ by the very construction of a tassel.) We will show, that
\be\label{c}
\|(1-\hat{p}_{s})\Psi\|\ =\ c(j(e_1),...,j(e_k))\|\Psi\|.
\ee 
The main point of this result is that the coefficient $c$ depends only on 
the labels assigned by the labelling $j$ to the edges. Of course
the coefficient is less then one, because the operator is
a projection and the trivial representation does appear
in the decomposition above. The new function $(1-\hat{p}_s)\Psi$ 
can be written as
\be
(1-\hat{p}_s)\Psi\ =\ \sum_n\Psi_n,
\ee  
where each $\Psi_n$ is again a spin-web state 
associated to a labelled web $(w_n,j_n)$ which contains 
a segment $s_1$ of the same properties as the segment $s_0$,
namely

$(i)$ $s_1$  is overlapped by $k$ edges,  $e'_1,...,e'_k$;

$(ii)$ the segments $e'_I$ overlapping $s_1$ are labelled
in the same way as the edges $e_I$, 
\be
j_n(e'_I)\ =\ j(e_I),\ \ \ I=1,...,k;
\ee

$(iii)$ the new segments $e'_{1+},..., e'_{k+}$ are holonomically
independent (where $e'_I=e'_{I+}\circ s_1\circ e'_{I-}$).

Therefore we can use the result again to conclude that
\be
\|(1-\hat{p}_{s_1}) (1-\hat{p}_{s_0})\Psi\|\ =\ c(j(e_1),...,j(e_k))^2
\|\Psi\|.
\ee
Finally, repeating the construction we find an infinite sequence of
segments $s_0, s_1,..., s_N,...$ such that 
\be
\|(1-\hat{p}_{s_N})...(1-\hat{p}_{s_0})\Psi\|\ =\ c(j(e_1),...,j(e_k))^{N+1}
\|\Psi\|.
\ee 
The sequence converges to $0$ as $N\rightarrow\infty$
hence indeed $\Psi$ is strongly degenerate.

Let us show now, that equation (\ref{c}) is true. The function 
$\Psi$ can be written as 
\be
\Psi\ =\ L_{A_1...A_k} {\D^{(j(e_1))}}^{A_1}_{B_1}(U_{e_1})...
{\D^{(j(e_k))}}^{A_k}_{B_k}(U_{e_k})R^{B_1...b_k},
\ee
where $L_{A_1...A_k},R^{B_1...B_k}\in \Cyl_w$ do not involve 
any of the holonomies $U_{e_1},...,U_{e_k}$. Then, the action
of the projection operator $\hat{p}_s$ on $\Psi$ is
\ba
\hat{p}_s\Psi \ =\ L_{A_1...A_k}{\D^{(j(e_1))}}^{A_1}_{C_1}(U_{e_{1+}})...
{\D^{(j(e_k))}}^{A_k}_{C_k}(U_{e_{k+}})c^{C_1...C_k}\nonumber\\
c_{D_1...D_k}
{\D^{(j(e_1))}}^{D_1}_{B_1}(U_{e_{1-}})...
{\D^{(j(e_{k-}))}}^{D_k}_{B_k}(U_{e_k})R^{B_1...B_k},
\ea
where
\be
c^{C_1...C_k}c_{D_1...D_k}\ =\ \int dU {\D^{(j(e_1))}}_{D_1}^{C_1}(U)...
{\D^{(j(e_k))}}_{D_k}^{C_k}(U).
\ee
The comparison of the norms gives the following result,
\be
\|\hat{p}_s\Psi\|^2\ =\ c^{C_1...C_k}\overline{c^{C'_1...C'_k}}
\delta_{C_1C'_1}...\delta_{C_kC'_k}c_{D_1...D_k}\overline{c_{D'_1...D'_k}}
\delta^{D_1D'_1}...\delta^{D_kD'_k}\|\Psi\|^2.
\ee

From the orthogonality of $\hat{p}_s\Psi$ and $(1-\hat{p}_s)\Psi$
we conclude (\ref{c}). 

It is easy to see, that as  mentioned, the web of Example in Sec. \ref{s3.1} 
belongs to the class of webs considered in this section. 
Indeed, each of the two edges contains a segment which is not overlapped 
by the other edge. Interestingly, even if we use the range of the web to
introduce more edges such that each regular segment is overlapped
by more than one edge, in all the examples we considered we found only 
webs belonging to the same class.
It is likely that for a semi-simple group this class contains all the webs.

What we have learnt from this section is that, as in the case of
the analytic category considered in \cite{ALMMT2}, the algebra of
the observables strongly commuting with the diffeomorphisms
is reducible to the sectors labelled by the ranges of the webs.   
In the case when a web $w$  is a graph, the corresponding
sector is $\H_w$. 
  
\section{The gravitational field operators}
\label{s5}
  
In this section for the group $G$ we take SU(2).
For the quantization of gravity there are two kinds of operators that play
a key role. The first kind consists of the so-called geometrical operators
\cite{RS2,AL4,L,TTL} corresponding respectively to lengths, areas and
volumes of curves, areas and volumes. The second kind consists of
constraint operators \cite{ALMMT2,QSDI}, most importantly the Hamiltonian
constraint. These have been defined in the analytical category so far. We
will now extend them to the smooth category.

\subsection{The 3-geometry operators}
\label{s5.1}

The regularizations of the  operators
representing geometric functionals can 
be easily modified to the case of the smooth curves. 
A problem one encounters then is that when a functional
depending on $E$ is given by an integral over a surface or a region 
in $\Sigma$ the resulting operator  often is not well defined
in any dense domain. This means that a given surface or region is just 
infinite in certain states of the system.

To begin with, consider a functional associated with an 
oriented 2-surface $S$ (see \cite{RS2,AL4} for the details)
\be
E^i_S\ =\ \int_S {1\over 2}E^{ia}\epsilon_{abc} dx^a\wedge dx^b, 
\ee
and the corresponding quantum operator
\be
\hat{E}^i_S\ =\ \int_S{1\over 2}{i\delta\over 
\delta A^i_a}\epsilon_{abc} dx^a\wedge dx^b. 
\ee
Let $\Phi\in \Cyl_w$ where $w$ is a web.   
`Generically', every edge $p$ of $w$ has only finite number of
isolated intersections with $S$ and only those
contribute to its action on $\Phi$. Generally, however,
that number is infinite. Every point $x$ of $e$ which is not 
contained in $S$ defines a segment $s$ of $e$ 
whose interior does not intersect $S$ whereas its ends
lie on $S$ or coincide with the ends of $e$.      
However, for each edge $e$ the number of such segments   
is at most countable. Split each such
segment into two pieces, oriented in such a way that they are outgoing from 
the surface and  
denote the resulting segments by $s_{e,1},...,s_{e,k},...$.
\begin{figure}
\epsfxsize=15cm
\centerline{\epsfbox{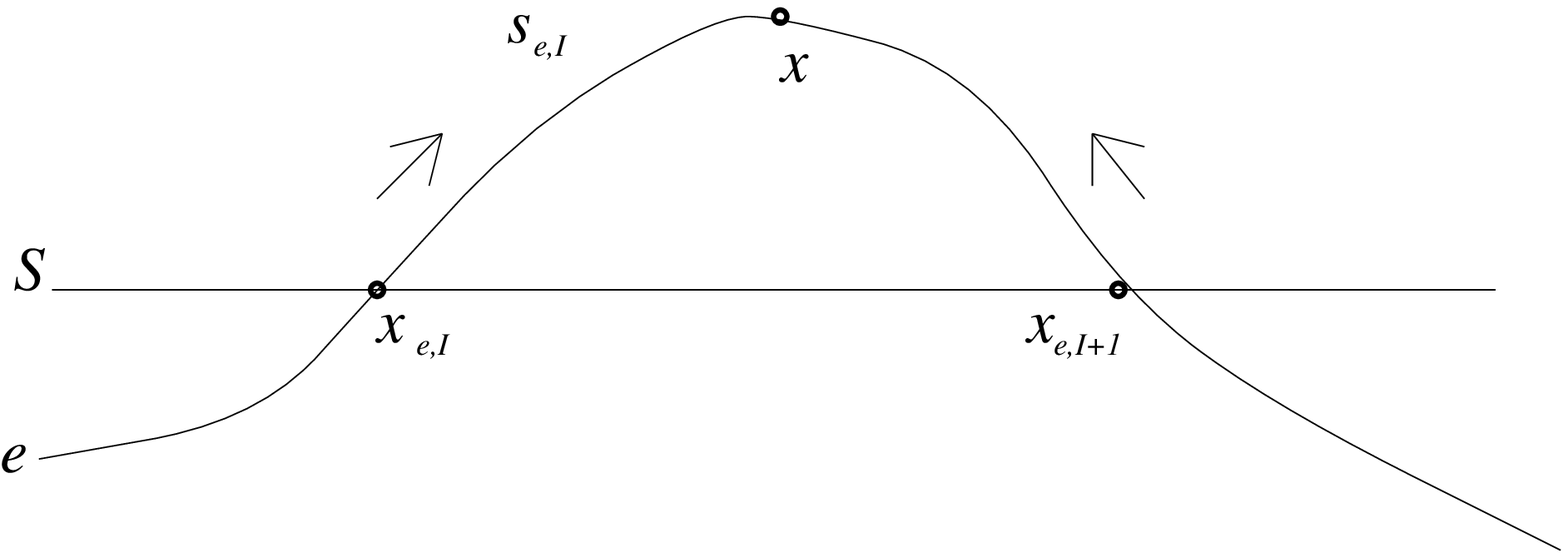}}
\vspace{2mm}
\end{figure}
Repeating the derivation of the first reference of \cite{AL4} we see that
for a cylindrical 
function $\Phi$ compatible with $w$, the only contribution
comes from those end points $x_{e,I}$ of the segments $s_{e,I}$ 
which are contained in $S$, according to the formula
\be
\hat{E}^i_S\Phi\ =\  {1\over 2}\sum_{s_{e,I}}
\kappa_S({s_{e,I}})J^i_{x_{e,I},s_{e,I}}\Phi,
\ee
where $\kappa_S({s_{e,I}})$ is $\pm 1$ if the segment intersects
$S$ and goes up/down (the above notation excludes segments which
overlap $S$ and contribute zero). The operators $J^i_{x_{e,I},s_{e,I}}$ 
were already defined in \cite{AL4} where it is understood that we replace 
graphs by webs. 

In the general case, for infinitely many intersections,
\be
\|\hat{E}^i_S \Phi\|\ =\ \infty.
\ee
Therefore, the domain of the operator is defined
by the generic webs which intersect the surface $S$
finitely many times. Using the same notation, for the area of 
$S$ we obtain
\be
\widehat{{\rm Area}}\Psi\ =\ 
{\Pl^2\over 2}\sum_{x_{e,I}}\sqrt{\big(\sum_{s}\kappa_S(s)
J^i_{x_{e,I},s}\big) \big(\sum_{s'}\kappa_S(s')
J^i_{x_{e,I},s'}\big)}\Psi,         
\ee
where $x_{e,I}$ runs through all the intersection points
of the segments $s_{e,I}$ with $S$ and 
$s$ and $s'$ run through all the segments $s_{e',I'}$
which intersect the same $x_{e,I}$. But again, the operator
may take the infinity. For instance, on a  Wilson loop function
\be
\Psi(A)\ =\ \Tr U_\alpha(A)
\ee
given by a non-self intersecting loop $\alpha$ which defines
infinitely many, transversally intersecting the surface $S$ segments 
$s_{e,I}$ defined above, we have
\be
\widehat{{\rm Area}}\Phi\ =\ \infty\Phi.
\ee

In the case of the volume of a region $R$ in $\Sigma$,  the `external' 
regularization of Rovelli, Smolin and DePietri \cite{RS2} does not appear
to be easily applicable to the smooth curves case. On the other
hand, the regularization of the second reference of \cite{AL4} proposed
for the volume operator can be extended to that case.
Then, the contribution is given by all the points 
of intersection of three edges of $w$ such that the
directions tangent to the edges are linearly
independent and the intersection point is contained in $R$ . 
The number of such generic intersection 
points is at most  countable.\footnote{Indeed, for any
given triple of edges $p,p',p''$ the generic  intersection points
are isolated from each other, so their set is 
countable; the number of triples is finite, determined by the number of
edges of the web.} 
The action of the resulting volume operator on a cylindrical function
compatible with the web $w$ reads   
\ba\label{volgamma}
\hat{V}_R\,\Psi\ &=&\ \k_o \sum_{v} \sqrt{|\q_v|}\, \Psi \quad
{\rm where} \nonumber\\ 
\q_v\, \Psi\ &=&\ {1\over 48}\e_{ijk}\sum_{s,s',s''}\e(s,s',s'')
J^i_{v,s}J^j_{v,s'}J^k_{v,s''}\, \Psi \, .  
\ea
where $v$ runs through all the intersection points
between the edges of $w$ and each of $s,s',s''$ runs through all the
segments intersecting at and are outgoing from $v$; the number
$\e(s,s',s'')$
equals $\pm 1, 0$ depending on the orientation
of the vectors tangent to $s,s',s''$ at $v$. \\
\\
Similar statements apply to the length operator $\hat{L}(c)$ \cite{TTL}
for any piecewise smooth curve $c$ since it 
uses the volume operator as a basic building block. \\
\\
Summarizing, for all three operators considered above, the infinity
arrises as a sum of a countable number of finite terms.
Therefore, if we exclude most of those points from the curve $c$, the 
surface $S$ or the region $R$ respectively then we will obtain
a finite length, area or volume. The process of gluing back
the removed points one after the other can be interpreted as 
extending the curve, surface or region to infinity. 
   
\subsection{Hamiltonian Constraint} 
\label{s5.2}

In Section \ref{s3.2} we have defined new solutions 
of the diffeomorphism constraint. They were, however, so far not shown
to be in the domain of the dual of the loop Hamiltonian constraint, as it
was defined in \cite{QSDI}. We observe below, that in fact the
action of the dual $\hat{H}'(N)$ of the Hamiltonian constraint can be 
naturally defined on them  and that it is free of the infinities 
emerging in the case of the above 3-geometry operators.

One can easily repeat the regularization of \cite{QSDI} of the Hamiltonian 
constraint, with appropriate modifications, to the case of a web with
an infinite number of intersection points among its defining curves.
Recall, that in the Hilbert space 
$\H$ of the cylindrical functions, one first defines  a regulator
depending Hamiltonian operator $H_\Delta$ where $\Delta$ denotes 
the regulator. 
Its domain is not dense in the smooth category any longer, in contrast 
to what happens in the analytical category.  
The reason for this is the same as in the case of the volume operator.
Explicitly, the regulated Hamiltonian operator acts on a spin-web state 
$|\Gamma\rangle$ in the following way 
\be \label{reg}
\widehat{H(N)}_\Delta |\Gamma\rangle \ =\ \sum_{v}N(v)\hat{h}_{\Delta(v)}
|\Gamma\rangle
\ee
where  $N$ is a laps function, $v$ runs through all the intersection 
points of the edges in a 
web and for each $v$ the operator $\hat{h}_{\Delta(v)}$ is well defined
for all the spin-web states. The number of the intersection points
is in general infinite, hence the total operator (\ref{reg}) is not well
defined on such states. (The operator $\hat{h}_{\Delta(v)}$
involves the volume operator in such a way that as in the volume
operator case, only the triples of generically intersecting
segments contribute). Remarkably however, the dual operator
${\widehat{H(N)}}_\Delta'$ is well defined for any dual spin-web
state, that is if $\Psi\in V_{w,j}$ and  $\Psi_1\in V_{w_1,j_1}$ for  
labelled webs $(w,j)$ and $(w_1,j_1)$, then  
$(\Psi|{\widehat{H(N)}_\Delta} \Psi_1)$ is finite. 

Finally, the  regulator free dual Hamiltonian
operator $\hat{H}'(N)$ is well defined on the diffeomorphism averaged dual 
spin web states $\langle\Gamma|$ provided that we restrict ourselves
to the space  $\T_{nondegfin}$ on which the diffeomorphism averaging
has been defined in Sec. \ref{s3.2}. More specifically,
whenever $\Gamma\in \T_{nondegfin}$ then for any lapse function 
the functional $\langle \Gamma|\hat{H}'(N)$ is a well defined
linear functional on the entire space  $\T_{nondegfin}$. \\

   
{\large Acknowledgements.}
The authors thank Abhay Ashtekar, John Baez and Steve Sawin for helpful 
discussions and correspondence. J.L. was supported by Alexander von 
Humboldt-Stiftung and the Polish Committee on Scientific Research
(KBN, grant no. 2 P03B 017 12).

\end{document}